# Raman Amplification of Matter Waves

Dominik Schneble,[*] Gretchen K. Campbell, Erik W. Streed, Micah Boyd, David E. Pritchard, and Wolfgang Ketterle

*MIT-Harvard Center for Ultracold Atoms, Research Laboratory of Electronics and Department of Physics,
Massachusetts Institute of Technology, Cambridge, MA 02139, USA*

(Dated: November 6, 2003)

We demonstrate a Raman amplifier for matter-waves, where the amplified atoms and the gain medium are in two different hyperfine states. This amplifier is based on a novel form of superradiance that arises from self-stimulated Raman scattering in a Bose-Einstein condensate.

With the realization of coherent, laser-like atoms in the form of Bose-Einstein condensates it has become possible to explore matter-wave amplification, a process in which the number of atoms in a quantum state is amplified due to bosonic stimulation. Stimulation has been observed in the formation of condensates [1, 2] and, more directly, has been used to realize coherent matter-wave amplifiers [3, 4] based on superradiant Rayleigh scattering [5–13] in which the atomic momentum of the gain medium and the amplified atoms differ by a photon recoil. In these cases the atoms remained in the same internal state, a fact that severely limited the performance of superradiant atom amplifiers since the amplified atoms were scattered out of the final state or served as a gain medium for higher-order processes (superradiant cascades [7]).

In this Letter we demonstrate a Raman atom amplifier in which the gain medium and the amplified atoms are in different internal states. Such a system has analogies to an optical laser in which different transitions are used for pumping and lasing, thus circumventing the above limitations. The gain mechanism for this amplifier is stimulated Raman scattering in a $\Lambda$-type atomic level structure which occurs in a superradiant way. This system also acts as a Stokes Raman laser for optical radiation.

The amplification scheme is similar to that explored in previous work on Rayleigh superradiance in Bose-Einstein condensates [7, 12] (cf. fig. 1 A). A linearly polarized laser beam with wavevector $\mathbf{k}$ is incident on a magnetically trapped, cigar-shaped condensate, perpendicular to its long axis. Each scattering event creates a scattered photon with momentum $\hbar(\mathbf{k} - \mathbf{q})$ and a recoiling atom with corresponding momentum $\hbar\mathbf{q}$. This scattering process is bosonically stimulated by atoms that are already present in the final state $\hbar\mathbf{q}$.

For Rayleigh scattering, the mode with the highest gain is the so-called endfire mode, in which the scattered photons propagate along the long axis of the condensate, while the scattered atoms recoil at an angle of $45°$ [7]. Rayleigh superradiance is strongest when the electric field vector of the incident beam is perpendicular to the long axis of the condensate and is suppressed when the field vector is parallel it [7]. This parallel configuration is the experimental situation considered in this Letter.

The angular distribution of spontaneously emitted light is dependent on its polarization. If we take the quantization axis $\hat{z}$ along the long axis of the condensate (which is also the direction of the bias field in the magnetic trap), the incident beam is $\pi$ polarized. If the emitted light is also $\pi$-polarized (Rayleigh scattering) its angular distribution is that of an oscillating dipole, $f_\pi(\theta) = (3/8\pi)\sin^2\theta$, where $\theta$ is the angle between $\hat{z}$ and the direction of propagation of the scattered light. The suppression of Rayleigh superradiance in this configuration is reflected in the fact that $f_\pi$ vanishes along $\hat{z}$, such that the endfire mode cannot be populated. If the emitted light is $\sigma$ polarized, the angular characteristics are that of a rotating dipole, $f_\sigma(\theta) = (3/16\pi)(1 + \cos^2\theta)$ and emission into the endfire mode is favored. The absorption of $\pi$ light followed by the emission of $\sigma$ light corresponds to a Raman transition in which the $z$ component of the atomic angular momentum changes by $\hbar$.

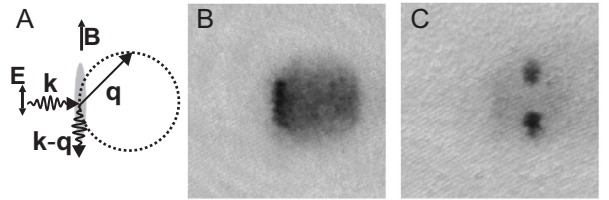

FIG. 1: Observation of superradiant Rayleigh scattering. (A) Experimental configuration. A laser beam (wave vector $\mathbf{k}$) is incident perpendicularly to the long axis of the condensate; its electric field vector $\mathbf{E}$ is parallel to it and the applied magnetic field $\mathbf{B}$. Each scattering event results in a recoiling atom (momentum $\hbar\mathbf{q}$) and a scattered photon (momentum $\hbar(\mathbf{k} - \mathbf{q})$). The recoiling atoms lie on a shell of radius $\hbar k$. (B) Spontaneous Rayleigh scattering. The absorption image shows a halo of atoms. The intensity of the beam was 1.0 mW/cm$^2$; the pulse duration was 1 ms. (C) Superradiant Raman scattering as observed for a beam intensity of 18 mW/cm$^2$ and a pulse duration of 100 $\mu$s (the original condensate was fully depleted after $\sim$ 10 $\mu$s). In both cases the field of view was 1.05 mm $\times$ 1.05 mm.

We now discuss the existence of superradiant gain for such a Raman process. For Rayleigh scattering an expression has been derived both semiclassically and quantum-mechanically, using descriptions based on either atomic or optical stimulation [7, 12, 14]:

$$\dot{N}_q = N_0 \, \Gamma_{sc} f(\theta) \, \Omega_q \, (N_q + 1) \equiv G(N_q + 1) \qquad (1)$$

Here, $N_0(N_q)$ is the number of atoms in the condensate (recoil mode), $\Gamma_{sc}$ is the rate for spontaneous Rayleigh scattering

(which is proportional to the beam intensity $I$), $f(\theta) = f_\pi(\theta)$ is the angular distribution of the spontaneously emitted light, and $\Omega_q \sim \lambda^2/A$ is the phase-matching solid angle for superradiance in an extended sample, where $\lambda$ is the wavelength of the scattered light and $A$ is the cross-sectional area perpendicular to $\hat{z}$. It is straightforward to generalize eq. 1 to the case of Raman scattering by interpreting $\Gamma_{sc}$ as the rate for spontaneous Raman scattering and replacing $f(\theta) = f_\sigma(\theta)$. The transition from spontaneous to superradiant (Rayleigh or Raman) scattering occurs when the gain exceeds the losses (see below), or $N_q > 1$. The angular distribution is then no longer the familiar dipole pattern $f(\theta)$, but becomes highly directional.

To demonstrate superradiant Raman scattering experimentally, a cigar-shaped $^{87}$Rb condensate was produced in a cloverleaf-type Ioffe-Pritchard magnetic trap with the procedure described in Ref. [12]. The condensate contained $N_0 = 10 \times 10^6$ atoms in the state $|1\rangle \equiv |5^2S_{1/2}, F = 1, m_F = -1\rangle$, where $F$ and $m_F$ are the quantum numbers for the total spin and its $\hat{z}$ component, and had Thomas-Fermi radii of 165 $\mu$m and 13.3 $\mu$m in the axial and radial directions, respectively. The trapped condensate was illuminated with a horizontal $\pi$ polarized laser beam for a variable duration $\tau$, and the magnetic trap was switched off immediately afterwards. The beam had a detuning of $\Delta/(2\pi) = -340$ MHz from the $5^2S_{1/2}, F = 1 \to 5^2P_{3/2}, F = 1$ transition at $\lambda = 780$ nm. The atomic momentum distribution was analyzed by imaging the atomic ensemble after 32 ms of ballistic expansion. Absorption images were obtained with a vertical probe beam after the atoms had been optically pumped to the state $5^2S_{1/2}, F = 2$.

Results obtained for light scattering in this situation are shown in fig. 1 B and C. For laser beam intensities $I$ of up to $I \sim 1$ mW/cm$^2$, a scattering halo consistent with a dipolar emission pattern was visible in the absorption images (cf. fig. 1 B), in agreement with earlier results [7] (for comparison, the threshold for Rayleigh superradiance was below $I = 100$ $\mu$W/cm$^2$ when the beam was polarized perpendicular to the long axis). However, when the intensity of the beam was increased to much larger values, the scattering became highly directional and two distinct peaks of recoiling atoms appeared into which the condensate was rapidly transferred (fig. 1 C). In contrast to the appearance of a fan pattern observed in Rayleigh superradiance due to higher-order scattering [7], the atoms remained localized in these two peaks, even when the light was left on. A Stern-Gerlach type analysis using ballistic expansion in a magnetic field gradient revealed that the atoms in the recoiling peaks had been transferred to the upper hyperfine state $|2\rangle \equiv |5^2S_{1/2}, F = 2, m_F = -2\rangle$, corresponding to the emission of $\sigma^+$ polarized light, Stokes-shifted by $-6.8$ GHz from the incident beam. Due to the change in the internal state the recoiling atoms were $\sim 6.5$ GHz out of resonance with the pump laser light, which explains the suppression of higher-order scattering. Previous experiments with sodium [7] used a large detuning which coincided with the ground state hyperfine splitting; thus the Raman scattered

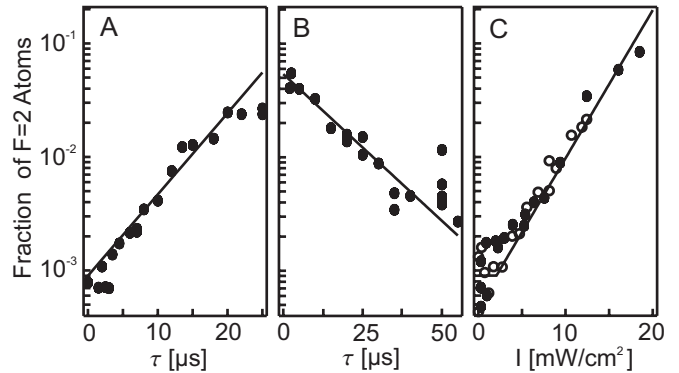

FIG. 2: Characterization of superradiant Raman scattering. The number of atoms in a single recoil peak is shown for for three different situations. A: Dependence on the pulse duration $\tau$, with constant beam intensity ($I = 7.6$ mW/cm$^2$). The solid line describes an exponential growth according to eq. (2) with $G - L = (1.7 \pm 0.1) \times 10^5$/s. B: Dependence on $\tau$ for a constant gain $G$ ($I\tau = 140$ mW/cm$^2$$\mu$s). The atom number decays exponentially with the rate $L = (6.2 \pm 0.5) \times 10^4$/s (solid line). C: Dependence on the laser intensity for a constant pulse duration ($\tau = 10$ $\mu$s). The solid line is calculated from eq. (2) with parameters $L$ and $G/I = 3.0 \times 10^4$ cm$^2$/mW s as obtained in A and B, predicting a threshold at 2.1 mW/cm$^2$. The open circles represent the measurement of Fig. 4, rescaled for the intensity of the electric field component along the long axis of the condensate.

atoms were in resonance with the pump light. It is most likely for this reason that superradiant Raman scattering has not been observed before.

The time evolution of the number $N_q$ of atoms in the recoil peaks in the perturbative regime $N_q \ll N_0$ (cf. fig. 2) showed exponential growth as predicted by eq. (1). In the presence of a loss mechanism with a rate $L$, the rate equation for $N_q \gg 1$ is modified to

$$\dot{N_q} = (G - L)N_q, \qquad (2)$$

To determine $L$, the condensate was illuminated with the beam for variable pulse durations $\tau$ while keeping the pulse area $I\tau \propto G\tau$ constant. In this case the number of recoiling Raman-scattered atoms should scale as $N_q(\tau) \propto \exp(-L\tau)$. The experimental result, shown in fig. 2 B, shows an exponential decay with a rate constant $L = 1/(16\mu s)$.

The loss rate $L$ observed here is a factor of 25 faster than the corresponding rate for superradiant Rayleigh scattering (i.e. the width of the two-photon Bragg resonance [7, 15]). To explain this difference we first note that, as in superradiant Rayleigh scattering, the term $N_0 N_q$ in eq. (1) describes an interference between the initial and final atomic states. This interference pattern constitutes a dynamic grating that diffracts the pump light into the endfire mode. As in the Rayleigh case, losses will occur when the overlap between the recoiling atoms and the condensate decreases. However, for Raman superradiance there is an important difference. The internal states are orthogonal, and therefore the spatial modulation is not in the atomic density but rather in the coherences (off-diagonal elements of the density matrix) [16] between the atomic states. Atomic coherences between different hyperfine states are sensitive to magnetic fields, leading to additional losses. In our case, there is a difference in the Zeeman shifts of states $|1\rangle$ ($m_F g_F = 1/2$) and $|2\rangle$ ($m_F g_F = -1$). In an inhomogeneous magnetic field the coherences evolve at locally varying frequencies, resulting in dephasing and thus in a decay of the grating. In a magnetic trap, the gravitational force is balanced by a magnetic field gradient. For the parameters of our trap, the variation of the magnetic field is 10 mG within the condensate, corresponding to an inhomogeneous broadening of the splitting between $|1\rangle$ and $|2\rangle$ of $2 \times 10^5$ s$^{-1}$. This simple estimate for $L$ agrees with the measured value to within a factor of 3.

The fast loss mechanism can also be explained by considering the population of the recoil state $\hbar\mathbf{q}$. In contrast to the atoms in the condensate the recoiling atoms are accelerated in the magnetic field gradient. After the dephasing time $L^{-1}$ their momentum has changed by $\hbar/d$, where $d$ is the vertical size of the condensate. This momentum transfer is comparable to the momentum uncertainty associated with $d$ and leads to a state distinct from the recoil state. This population loss in $\hbar\mathbf{q}$ could be decreased by avoiding magnetic field gradients, e.g. in an optical trap or by using free falling condensates.

The dependence of the atom number in the recoil peaks on the intensity of the laser beam is shown in fig. 2 C. The atom number increased exponentially according to $N_q(\tau) \propto \exp(G - L)\tau$ above threshold, with parameters that are consistent with those found in the measurements shown in fig. 2 A and B. The experimentally determined gain $G$ agrees with the value predicted by eq. (1) ($G_{th} = 1 \times 10^6$/s for a beam intensity of 7.6 mW/cm$^2$ [17]) to within a factor of 4.

Raman superradiance can be considered as amplification of the first, spontaneous scattering event into the endfire mode. However, if a coherent input is applied to the system this input will get amplified as well. In analogy to the work on Rayleigh scattering [3, 4] we have used superradiant Raman scattering as a mechanism for the amplification of matter-waves, as illustrated in fig. 3. To prepare the input, a short laser pulse was first applied to the trapped condensate, preparing a small cloud of $N_q$ recoiling atoms in state $|2\rangle$. A second, identical pulse was then applied after a wait time $\tau'$. This led to a significant increase of the atom number to an output value $N'_q = \beta N_q$, where $\beta$ is the amplification factor. Values up to $\beta = 10$ were observed (fig.3), which is a clear signature for Raman amplification in state $|2\rangle$.

In agreement with the loss mechanism discussed above, the matter-wave amplification factor $\beta$ was found to depend on the wait time $\tau'$ between the pulses. During this time, the input atoms were accelerated out of the narrow bandwidth of the amplifying pulse. The observed amplification factor (cf. fig. 3 D) decayed with a rate of $2.2 \times 10^6$/s to a value of $\beta = 2$ in which case the number of atoms in the output was simply the sum of two independent superradiant pulses. This decay rate is in good agreement with the estimate for $L$ given above and agrees with the result of fig. 2 B to within a factor of 3.5. Both rates should agree according to eq. (2), but we note that our simple model does not account for the depletion of the condensate and optical propagation effects (note that for the measurement shown in fig. 2 B the pump light was on, while it was off between the two pulses).

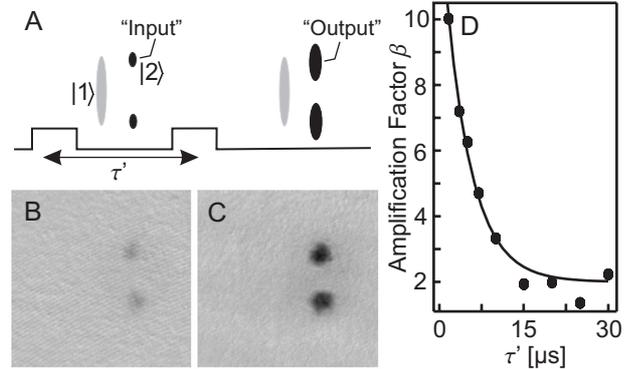

FIG. 3: (A): Realization of a Raman amplifier for matter waves. The first laser pulse prepares a cloud of $N_q$ recoiling atoms in state $|2\rangle$ (input). After a wait period $\tau'$, these atoms are amplified by a second, identical pulse to a number $N'_q = \beta N_q$ (output), where $\beta$ is the amplification factor. (B) and (C): Input and output ($\beta = 6$) for a pulse duration $\tau = 1.65$ $\mu$s and a wait time $\tau' = 5$ $\mu$s. The intensity of the laser beam was 60 mW/cm$^2$. The absorption images were taken without optical pumping from $|1\rangle$ to $|2\rangle$ and therefore do not show the condensate in the center of the field of view. (D): Dependence of the amplification factor $\beta$ on the wait time $\tau'$. The line represents an exponential decay with a decay rate of $(2.2 \pm 0.2) \times 10^5$/s.

Superradiant Rayleigh and Raman scattering can be regarded as two modes of an atom laser. Just as in a multimode optical laser one mode can be selected by introducing mode-selective elements, the relative gain of the two matter-wave modes can be controlled by the polarization of the incident light, as is shown in fig. 4. When the polarization was rotated from parallel to the long axis of the condensate to perpendicular, the three-peaked Raman pattern of recoiling atoms (A) changed into the X-shaped pattern (B) characteristic for the Kapitza-Dirac regime of superradiant Rayleigh scattering [12, 13]. The polarization angle dependence is shown in (C). The transition was almost symmetric about an angle of 45°, around which both types of superradiance were active. Note however that this symmetry is not universal but depends on the particular choice of parameters. The thresholds (loss rates) for the two kinds of superradiance differ by a factor of 25, and the transition matrix element for Raman scattering is 6.5 times smaller than that for Rayleigh scattering at our detuning. The fact that the superradiant rates at the optimum angles were almost the same is due to the fact that at short pulses, the superradiant Rayleigh gain is suppressed due to the presence of backward peaks [12]. In the present case, which is far above threshold and near the end of the short-pulse regime [12], this effect resulted in a gain suppression by a factor $\sim 6.5$ for Rayleigh superradiance.

The Raman transition from $|1\rangle$ to $|2\rangle$ involves the absorp-tion of $\pi$ light and emission of $\sigma^+$ light. One can therefore regard the angular dependence of superradiant Raman scattering as a dependence on the intensity of the $\pi$ component of the incident light, which scales as $\cos^2 \alpha$, where $\alpha$ is the angle between the electric field vector and the axis. Indeed, the rescaled data of fig. 4 nicely matches the other data shown in fig. 2 (C) for which the total intensity of the light was varied.

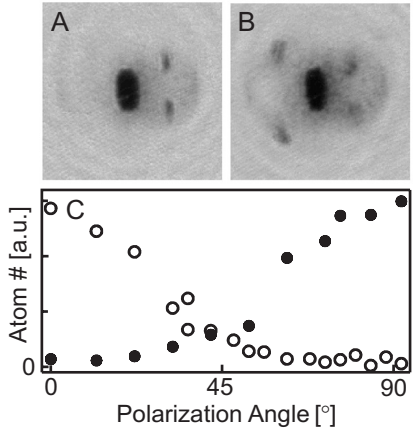

FIG. 4: Tuning between superradiant modes. (A) Raman superradiance (B) Rayleigh superradiance (Kapitza-Dirac regime) for polarization parallel and orthogonal to the condensate axis, respectively. (C) Variation of the atom number in the Raman peaks (open circles) and in the Rayleigh backward peaks (filled circles) with the polarization angle. The angle had a systematic uncertainty of $\sim 10°$ from the orientation of the condensate with respect to the axis of the magnetic trap. The measurements were done for a laser intensity of 12 mW/cm$^2$ and a pulse duration of 10 $\mu$s.

The comparison of these two amplification modes also illustrates that the observed Raman superradiance does not have a short-pulse (or Kapitza-Dirac) regime, unlike the Rayleigh process for which endfire mode photons are scattered back into the laser beam [12]. This process violates energy conservation by the recoil frequency and therefore can happen only at times shorter or comparable to the inverse recoil frequency. For the observed Raman scattering, the energy mismatch (divided by $h$) is $\sim 6.5$ GHz, corresponding to sub-ns time scales on which backscattering might happen. Similarly, the absence of higher-order forward peaks reflects the fact that the laser light is too far out of resonance with the atoms in the recoil state to drive further Raman transitions (e.g., back to $F = 1$) on the experimental time scale.

Finally, we mention that superradiant Raman scattering is also possible between sublevels of the same hyperfine state. For a detuning of $\Delta/(2\pi) = -140$ MHz, the transition matrix elements for Raman transitions to $5\,^2S_{1/2}, F = 2, m_F = -2$ and to $5\,^2S_{1/2}, F = 1, m_F = 0$ are almost the same. Indeed we observed strong Raman superradiance with a $F = 1, m_F = 0$ first-order peak and a $F = 1, m_F = 1$ second order peak (this was observed also for $-340$ MHz detuning, but was much less pronounced). Further peaks (orders higher than 2 or backward peaks) were not observed. For longer pulse length, the $F = 1$ peaks disappeared, unlike the $F = 2$ peaks, since subsequent spontaneous Rayleigh scattering is much stronger for $F = 1$ atoms due to the much smaller detuning.

In summary, the long coherence time of condensates has allowed us to study new phenomena in the interaction of an atomic cloud with a single strong laser beam. The cloud can spontaneously create not only density gratings [7] and phase gratings [12], but also periodic modulations of coherences. Whereas an atomic density grating is accompanied by a standing light wave formed by the pump light and the endfire mode, the coherence grating is driven by an optical field with a polarization grating formed by two beams of orthogonal polarization. The exponential growth of coherence gratings provides a mechanism to amplify matter waves in an internal state different from the condensate.

We would like to thank J. M. Vogels and J. Steinhauer for fruitful discussions, and A. E. Leanhardt for a critical reading of the manuscript. We acknowledge financial support by NSF and the MIT-Harvard Center for Ultracold Atoms.

intermediate state to the final state, resulting in a dependency $\Gamma_{sc} \propto (dd'/\Delta)^2$. In this expression $d$ and $d'$ are the dipole matrix elements for the Raman transition. The couplings to the various states $i$ of the excited-state manifold were included by performing a coherent summation, $\Gamma_{sc} \propto (\sum d_i d'_i/\Delta_i)^2$.